\documentclass[12]{article}
\usepackage{amsmath,color,xcolor}
\usepackage{graphics,graphicx,amsbsy,amssymb}

\newcommand{\JJ}{{\boldmath \mbox{$J$}}}

\newcommand{\uu}{{\boldmath \mbox{$u$}}}

\newcommand{\XX}{{\boldmath \mbox{$X$}}}

\newcommand{\rr}{{\boldmath \mbox{$r$}}}

\newcommand{\vv}{{\boldmath \mbox{$v$}}}

\newcommand{\ddelta}{{\boldmath\mbox{$\delta$}}}

\newcommand{\nnu}{{\boldmath\mbox{$\nu$}}}
\newcommand{\ssigma}{{\boldmath\mbox{$\sigma$}}}

\newcommand{\dd}{\mbox{d}}

\newlength{\defbaselineskip}
\newcommand{\setlinespacing}[1]%
           {\setlength{\baselineskip}{#1 \defbaselineskip}}

\begin{document}








\title{\textbf{Landau damping in   the  multiscale    Vlasov  theory}}
\author{Miroslav Grmela$^{1}$\footnote{corresponding author: e-mail:
miroslav.grmela@polymtl.ca},$\;$ Michal Pavelka $^{1,2}$\vspace {0.5cm}\\
$^1$ \'{E}cole Polytechnique de Montr\'{e}al,\\
  C.P.6079 suc. Centre-ville,
 Montr\'{e}al, H3C 3A7,  Qu\'{e}bec, Canada \\
$^2$ Mathematical Institute, Faculty of Mathematics, Charles University,\\ Prague,
 Sokolovsk\'{a} 83, 18675 Prague, Czech Republic
 }
 \maketitle
 \date{}

\begin{abstract}

Vlasov  kinetic theory is extended by adopting an extra one particle distribution function as an  additional state variable characterizing the micro-turbulence internal structure. The extended Vlasov equation keeps the  reversibility, the Hamiltonian structure, and the entropy conservation  of the original Vlasov equation. In the setting of the extended Vlasov theory we then   argue  that  the Fokker-Planck type damping in the velocity dependence of the extra  distribution function   induces the Landau  damping. The same type of extension is  made also in the setting of fluid mechanics.

\end{abstract}

\section{Introduction}\label{11}

The physical system under investigation is a gas composed of particles interacting via  long range forces. We call this gas a Vlasov gas rather than  plasma in order to emphasize that the interaction forces that we consider in this paper are not associated with
any type of field (e.g. the electromagnetic field) that has its own dynamics (e.g. Maxwell's electromagnetic field theory).
Because of  the presence of long range interactions,  we expect   to see in the Vlasov  gas a collective behavior and thus  large pair and even higher order correlations. In the position space, the correlations manifest themselves in  the formation of fine scale spatial structures, in the velocity space, in the micro-turbulence.
It is thus  surprising that the Vlasov kinetic theory, with only one particle distribution function serving as the state variable and with the mean-field consideration of long range interactions, has proved to provide a good theoretical description of the behavior observed in experiments. In this paper we put into focus one particular experimental observation that became known as Landau damping. It is the time irreversible evolution of the one particle distribution function toward spatial homogeneity. Before being experimentally observed \cite{expLandau}, this behavior was predicted \cite{Landau} from an analysis of solutions of the Vlasov kinetic equation. Landau damping is particularly surprising since the Vlasov kinetic equation is time reversible. How can the  reversible time evolution equation have time irreversible solutions? The recent  mathematical analysis in \cite{VillaniV}, extending Landau's investigation of the linearized Vlasov equation to the complete nonlinear setting, points out to the analogy, both in the physics and the mathematics involved, with the general conundrum of statistical mechanics, namely with the question of  how does the reversible Hamiltonian time evolution  of an isolated system of $10^{23}$ particles imply the observed time irreversible approach to thermodynamic equilibrium states. Inspired by this analogy, we investigate in this paper  the Landau damping by using some of the methods developed originally in statistical mechanics.

Following Boltzmann,  the tendency of  solutions of the microscopic time evolution equations to increase their  irregularity is mathematically expressed in the emergence of dissipative and irreversible time evolution of appropriately chosen mesoscopic state variables. This strategy can be seen as a way to regularize the complex microscopic time evolution. The loss of regularity and the time irreversible weak convergence seen in the microscopic time evolution becomes after regularization an explicitly visible regular property of the time evolution of  appropriately chosen mesoscopic state variables in which many microscopic details  are ignored. The weak convergence is transformed in the regularization into the strong convergence.
In the example  of the  Boltzmann gas (gas particles interact only via very short range repulsive forces), the particle trajectories become very complex (their irregularity increases) due to  strong interactions experienced by the gas particles during collisions. Boltzmann's regularization of the Boltzmann  gas dynamics consists in choosing only one particle distribution function as the state variable and taking into account the influence of collisions in  the  Boltzmann collision term  that is dissipative and time irreversible. Solutions to the Boltzmann equation are  regular from the mathematical point of view and show approach  to the Maxwell distribution in velocities and to the spatial homogeneity, both observed in experiments.

One possible way to regularize the Vlasov equation is to add to its right hand side a self-diffusion term that  drives solutions to the spatial homogeneity. For such regularized Vlasov equation the Landau damping becomes  a regular and explicitly visible property of its solution. In order to see the  physics that is behind the added self-diffusion  term, we
step out of the confinement of the Vlasov one particle kinetic theory and  formulate (in Section \ref{3}) the time evolution of the Vlasov gas
on a different, more detailed,  level. Two questions arise: first, how do we formulate such extended Vlasov theory, and second, how does the Landau damping present itself in solutions of the governing equations of the extended theory.

The paper is organized as follows. In Section \ref{1} we present a general nondissipative one particle kinetic equation  and recall three properties of their solutions (reversibility, Hamiltonian structure, and entropy conservation). A special case, corresponding to the Boltzmann gas,  is reviewed   in Section \ref{2}. We introduce the Boltzmann regularization (based on considering collisions as the  source of dissipation). In Section \ref{3}, we construct extended Vlasov  kinetic equation (whose solutions still possess the three properties identified in Section \ref{1} for one particle kinetic equations), regularize them (based on considering the decay of micro-turbulence as the source of dissipation) and show then how this dissipation induces  the Landau damping in the one particle distribution function. In Section \ref{4} we make the same type of  investigation of the Vlasov gas but with the level of fluid mechanics replacing the level of kinetic theory.

\section{Nondissipative one particle kinetic theory}\label{1}

Let $f(\rr,\vv)$  be one particle distribution function,   $\rr$ and $\vv$  denote the position coordinate and momentum of one particle. Its time evolution is governed by
\begin{equation}\label{KE}
\frac{\partial f(\rr,\vv)}{\partial t}=L^{(f)}E^{(f)}_{f(\rr,\vv)}
\end{equation}
where  $E(f)$ is a real valued function of $f$ having the physical meaning of energy, $E_f$ denotes derivative of $E$ with respect to $f$,  $L^{(f)}$ is the Poisson bivector expressing mathematically the kinematics of $f$. It is  given   by associating it with the Poisson bracket
\begin{eqnarray}\label{L}
&&\int d\rr\int d\vv A_{f(\rr,\vv)}L^{(f)}B_{f(\rr,\vv)}=\{A,B\}^{(f)}\nonumber \\
&&\{A,B\}^{(f)}=\int d\rr\int d\vv f\left[\frac{\partial A_{f}}{\partial \rr}\frac{\partial B_{f}}{\partial \vv}-\frac{\partial B_{f}}{\partial \rr}\frac{\partial A_{f}}{\partial \vv}\right]
\end{eqnarray}
where $A$ and $B$ are real valued as sufficiently regular functions of $f$. We recall that $\{A,B\}^{(f)}$ given in (\ref{L}) is a Poisson bracket if  $\{A,B\}^{(f)}=-\{B,A\}^{(f)}$ (which is manifestly visible)  and the Jacobi identity $\{A,\{B,C\}^{(f)}\}^{(f)}\\+\{B,\{C,A\}^{(f)}\}^{(f)}+\{C,\{A,B\}^{(f)}\}^{(f)}=0$ holds (which can either be verified directly or by associating it with the Lie group of canonical transformations of $(\rr,\vv)$ - see \cite{MW}). As for the boundary conditions, we require that all the integrals over the boundary arising in by parts integration disappear. Written explicitly, the kinetic equation (\ref{KE}) takes the form
\begin{equation}\label{KEexpl}
\frac{\partial f}{\partial t}= -\frac{\partial}{\partial \rr}\left(f\frac{E^{(f)}_f}{\partial\vv}\right)+\frac{\partial}{\partial \vv}\left(f\frac{E^{(f)}_f}{\partial\rr}\right)
\end{equation}

From the mathematical point of view, (\ref{KE}) is a general formulation of the Hamiltonian dynamics, i.e.  a covector constructed as a gradient of a potential (having the physical meaning of energy) is transformed into a vector (serving as a vector field generating the time evolution) by a Poisson bivector. From the physical point of view, (\ref{KE}) expresses the mechanics. It may not be the classical canonical mechanics (in which the role of $f$ is played by $(\rr,\vv)$ and $L^{(rv)}=\left(\begin{array}{cc}0&1\\-1&0\end{array}\right))$ but still it is the time evolution possessing the physically essential features of mechanics. The noncanonical formulation (\ref{KE}) of mechanics has been introduced by Arnold \cite{Arnoldfluid} in the context of fluid mechanics (putting the Euler equation, that has already been proven to represent the Hamiltonian dynamics by Clebsch in \cite{Clebsch}, into the form (\ref{KE})). The Poisson bracket (\ref{L}) has been introduced in \cite{MW}.

Now we turn to important qualitative properties of solutions to (\ref{KE}).
It follows directly from the general formulation  (\ref{KE}) that solutions to the kinetic equation (\ref{KE}) possess the following three properties:
\\

\textbf{\textit{Conservation of the energy}}

The skewsymmetry of $L^{(f)}$ implies immediately the conservation of the energy $E^{(f)}$
\begin{equation}\label{consE}
\frac{dE^{(f)}}{dt}=0,
\end{equation}
Indeed, $\frac{dE^{(f)}}{dt}=\{E^{(f)},E^{(f)}\}^{(f)}=0$. The energy conservation is indeed a cornerstone of the physics involved in  mechanics.
\\

\textbf{\textit{Conservation of entropy}}

Another conservation appears as a consequence of the degeneracy of $L^{(f)}$. The degeneracy is defined as follows.  If there exists a potential $C^{(f)}(f)$ (in addition to the potential $C^{(f)}(f) = const.$) for which $\{A,C\}^{(f)}=0$ for all $A(f)$ then $L^{(f)}$ is called degenerate and the potential $C^{(f)}f)$ is called Casimir of the Poissson bracket $\{A,B\}^{(f)}$. Since $\frac{dC^{(f)}}{dt}=\{C^{(f)},E^{(f)}\}^{(f)}$, we have then the conservation law
\begin{equation}\label{consS}
\frac{d C^{(f)}}{dt}=0
\end{equation}

It can be directly verified that  potentials of the form $C^{(f)}(f)$\\$=\int d\rr\int d\vv c(f(\rr,\vv))$, where $c$ is a sufficiently regular function  $c:\mathbb{R}\rightarrow \mathbb{R}$, are all  Casimirs of the Poisson bracket (\ref{L}). Equation (\ref{consS}) represents thus an infinite number of conservation laws that supplement   the energy conservation law (\ref{consE}). We note that in the completely microscopic classical  mechanics of particles these additional conservation laws are absent
since the Poisson bivector arising there  is nondegenerate. The essential difference between the microscopic particle mechanics and mesoscopic mechanics (as e.g. the kinetic theory represented by (\ref{KE})) is that the former keeps all the details while the latter  ignores some of them. The ignored details have not however disappeared from the
time evolution of the mesoscopic state variables ( i.e. the state variables   that we have decided to keep;   in kinetic theory we keep  the one particle distribution function). How shall we express their influence? It is thermodynamics that answers this question.  The influence of the ignored  details on   the time evolution of the mesoscopic state variables is expressed in the time evolution of a new potential, called entropy (we shall use the symbol $S$ to denote it). The entropy $S$ tends to its maximum allowed by constraints. In the case of the nondissipative mesoscopic dynamics that we discuss in this section, the entropy $S$ is a potential that remains unchanged. In the nondissipative mesoscopic dynamics the ignored details manifest themselves in the emergence of a companion conservation law.  Since the Casimirs $C$ are conserved (see (\ref{consS})),  we choose the entropy to be one of the Casimirs $C(f)$. To continue and select one  among the infinite number of Casimirs  requires an investigation of  the dissipative dynamics (in Sections \ref{2} and \ref{3}).
\\

\textit{\textbf{Time reversibility}}

So far, we have made no restriction on the choice of the energy $E^{(f)}(f)$. Now, we  make one. We assume that the energy is invariant with respect to  the transformation $\vv\rightarrow -\vv$. If this is the case then we see immediately that (\ref{KE})
 is time reversible in the sense that the transformation $t\rightarrow -t$ is fully compensated by the transformation  $\vv\rightarrow -\vv$ (i.e.
the trajectory $f(\rr,\vv,t)$ corresponding to the initial condition $f_0(\rr,\vv)$ is the same as the trajectory $f(\rr,-\vv,-t)$ corresponding to the initial condition $f_0(\rr,-\vv)$).
\\

\section{Boltzmann kinetic theory}\label{2}

In this section we recall several well known results about the Boltzmann kinetic equation. Our objective is to prepare the setting for discussing  the Vlasov kinetic equation in Section \ref{3} and also to highlight  similarities and  differences between the Boltzmann and the Vlasov   kinetic equations.

\subsection{Nondissipative Boltzmann equation}

The physical system under investigation in this section is the Boltzmann gas (we use the superscript  $(fB)$  to denote  the quantities that arise in this investigation).
The energy $E^{(fB)}(f)$ is given by
\begin{equation}\label{energyB}
E^{(fB)}(f)=E^{(fB)}_{(kin)}(f)+E^{(fB)}_{(pot)}(f)
\end{equation}
where
\begin{equation}\label{enkin}
E^{(fB)}_{(kin)}(f)=\int d\rr\int d\vv f(\rr,\vv)\frac{\vv^2}{2m}
\end{equation}
is the kinetic energy, $m$ denotes the mass of one gas particle,   and
\begin{equation}\label{enpot}
E^{(fB)}_{(pot)}(f)=\frac{1}{2}\int d\rr_1\int d\vv_1 \int d\rr_2\int d\vv_2 f_2(f;\rr_1,\vv_1,\rr_2,\vv_2)V^{(fB)}(|\rr_2-\rr_1|)
\end{equation}
The potential $V^{(fB)}$ is a very short-range hard-core type two-particle repulsive potential, $f_2(f;\rr_1,\vv_1,\rr_2,\vv_2)$ is the two particle distribution function expressed in terms of the one particle distribution function $f$.
The Boltzmann gas particles are free, they are not subjected to any force,  except when two particles become very close. In such case, the two particles are subjected to a large  repulsive force. The kinetic equation governing the time evolution of the Boltzmann gas is thus (\ref{KE}) with the energy (\ref{energyB}), i.e.
\begin{equation}\label{BEfull}
\frac{\partial f}{\partial t}=-\frac{\partial}{\partial \rr}\left(f\frac{E^{(fB)}_f}{\partial\vv}\right)+\frac{\partial}{\partial \vv}\left(f\frac{E^{(fB)}_f}{\partial\rr}\right)
\end{equation}

What can we say about solutions to (\ref{BEfull})? First, we note that
the four properties
\begin{eqnarray}\label{BEfth}
&&\frac{dE^{(fB)}}{dt}=0\nonumber \\
&&\frac{dC^{(f)}}{dt}=0\nonumber \\
&&\frac{dN^{(B)}}{dt}=0\nonumber \\
&& time\,\,reversibility
\end{eqnarray}
proven in Section \ref{1} hold provided $f_2(f;\rr_1,\vv_1,\rr_2,\vv_2)$ is sufficiently regular and invariant with respect to $(\vv_1,\vv_2)\rightarrow (-\vv_1,-\vv_2) $ function and $C$ is a Casimir of the bracket (\ref{L}). The potential $N^{(B)}(f)=\int d\rr\int d\vv f(\rr,\vv)$ is a Casimir that has the physical interpretation of the number of moles.

Regarding more detailed properties of solutions to (\ref{BEfull}), we recall that  experimental observations of the Boltzmann gas show that, in the absence of external forces, the Boltzmann gas approaches (as $t\rightarrow\infty$) a distribution function $f_{eq}(\rr,\vv)$ that is independent of $\rr$ and Maxwellian (i.e. proportional to $e^{-\frac{\vv^2}{2mk_BT}}$, where $k_B$ is the Boltzmann constant and $T$ is the equilibrium temperature) in the dependence on $\vv$. Is this experimental observation seen in solutions to (\ref{BEfull})?

To  begin with  answering this question,  we  first recall
(see \cite{VillaniV}) that in the absence of
the potential energy (\ref{enpot})  in the energy (\ref{energyB}),  solutions to (\ref{BEfull}) converge weakly in large time to a spatially homogeneous distribution that is equal to the space average of the initial distribution function $f_0(\rr,\vv)$. We expect that this property will  still hold for the full kinetic equation (\ref{BEfull}) but, to the best of our knowledge, such property of solutions to (\ref{BEfull}), as well as the Maxwellian dependence on $\vv$ in the asymptotic solutions to (\ref{BEfull}),   has not been rigorously proven.
There are many results coming from  direct simulations of  systems composed of (up to $10^5$ ) particles interacting via  potential of the type (\ref{enpot}) but such results, while addressing the time evolution of the  Boltzmann gas, are not results about solutions to the kinetic equation (\ref{BEfull}). Nevertheless, we may assume that solutions to (\ref{BEfull}) show time irreversible tendency towards spatial homogeneity and Maxwell distribution in $\vv$. This tendency is not seen in the properties (\ref{BEfth}).

In the further investigation of solutions to (\ref{BEfull}) we shall  follow Boltzmann and make the time irreversible tendency of solutions to (\ref{BEfull}) manifestly visible by modifying Eq.(\ref{BEfull}) into a new equation  whose solutions  have the properties (\ref{BEfth}) except that the last line in (\ref{BEfth}) is absent and  in the  second line the Casimir $C^{(f)}(f)$ is a one specific  Casimir called the Boltzmann entropy and the equality is changed into an inequality. It is this inequality and the time irreversible nature of the equation that brings into the visibility the time irreversible tendency of solutions to (\ref{BEfull}).
The modified (regularized) kinetic equation (\ref{BEfull})  is then  called a Boltzmann equation. The transformation of  (\ref{BEfull}) into the Boltzmann equation is in fact  a first step in the investigation of solutions to  (\ref{BEfull}). The Boltzmann equation is simpler than  (\ref{BEfull}), it is easier to solve it, and its solutions  are  expected (the expectation is based on
the physical insight leading   from (\ref{BEfull} to the Boltzmann equation)  to approximate well solutions to  (\ref{BEfull}). This is the Boltzmann strategy (recalled below in the context of the Boltzmann gas) that we shall also follow in the context of the Vlasov gas in Section \ref{3}.

\subsection{Regularization}\label{Breg}

The physical insight on which the Boltzmann regularization of (\ref{BEfull})  is based is that the very rapid and very large  changes in momenta of gas particles during collisions are responsible for the spatial homogenization and also for the Maxwell distribution in velocities seen in the one particle distribution function. Details of the trajectories during collisions are ignored in the Boltzmann analysis.

The first step in the Boltzmann modification is a complete  omission of the hard-core repulsive potential energy (\ref{enpot}). Such omission is justified if  the kinematics is modified in such a way that the particles are prevented to be close enough to make  the potential energy (\ref{enpot})  different from zero. This is easy to do if we regard the particles as point particles so that, even without changing kinematics, the potential energy (\ref{enpot}) completely  disappears.
The total energy of the Boltzmann gas is thus only the kinetic energy of the gas particles. But even with this idealized particles, we still have to admit that collisions  occur and that they  influence the behavior of the gas. Having  no mechanical basis to express their influence (recall that the second term on the right hand side of (\ref{KE}) is now absent) we turn to chemical kinetics. Collisions are  seen in the Boltzmann analysis as  chemical reactions. The momentum $\vv$ serves as the variable parametrizing the species. The chemical reaction representing the binary collision is a reaction  in which two species, one with the parameter  $\vv$ and the other with the parameter  $\vv_1$,    interact. The outcome of the reaction are two new species, one with the parameters   $\vv'$ and  $\vv'_1 $. The mechanical nature of this "chemical reaction" is expressed in two constraints,
\begin{eqnarray}\label{constcoll}
\vv^2+\vv^2_1&=&\vv'^2+\vv'^2_1\nonumber \\
\vv+\vv_1&=&\vv'+\vv'_1
\end{eqnarray}
having the physical meaning of  the energy  the momentum conservations in binary collisions. With this viewpoint of collisions,
the second term on the right hand side of (\ref{BEfull})  becomes  replaced with  the so called Boltzmann collision term $\left(\frac{\partial f}{\partial t}\right)_{diss}$. With such replacement, the kinetic equation (\ref{KE}) turns into the Boltzmann kinetic equation
\begin{equation}\label{BE}
\frac{\partial f}{\partial t}= -\frac{\partial}{\partial \rr}\left(f\frac{\vv}{m}\right)+\left(\frac{\partial f}{\partial t}\right)_{diss}
\end{equation}

For the mathematical expression of $\left(\frac{\partial f}{\partial t}\right)_{diss}$ we turn now to the mass action law in chemical kinetics
 (see \cite{Grchem}):
\begin{equation}\label{collW}
\left(\frac{\partial f}{\partial t}\right)_{diss}=\left[\Xi^{(B)}_{f^*}(f,f^*)\right]_{f^*=S^{(B)}_f}
\end{equation}
where
$\Xi^{(B)}(f,f^*)$. called a dissipation potential, is given by
\begin{eqnarray}\label{Xi}
\Xi^{(B)}(f,f^*)&=&\int d\rr\int d\vv\int d\vv'\int d\vv_1\int d\vv'_1W^{(B)}(f,\vv,\vv',\vv_1\vv'_1)\nonumber \\
&&\times\left[e^{X^{(B)}}+e^{-X^{(B)}}-2\right]
\end{eqnarray}
the thermodynamic force $X^{(B)}$ (called affinity in the context of chemical kinetics)  by
\begin{equation}\label{X}
X^{(B)}(f^*)=f^*(\rr,\vv')+f^*(\rr,\vv'_1)-f^*(\rr,\vv)-f^*(\rr,\vv_1)
\end{equation}
$W^{(B)}$,  playing  the role of the rate coefficients,  equals zero except when  (\ref{constcoll}) holds and then it is positive,
and $S^{(B)}(f)$
is a Casimir (see (\ref{consS})). The mechanical origin of binary collisions is retained  in (\ref{collW}) only in the constraints (\ref{constcoll} expressing mechanical conservation laws, and, indirectly, in the choice of $S^{(B)}(f)$ that we shall discuss below.
It can directly be verified that the  collision term  is invariant with respect to the transformation $\vv\rightarrow -\vv$  and thus it is the term that  breaks  the time reversibility of the equation (\ref{BEfull}).

\subsection{Properties of solutions}\label{PSBE}

Now we turn to solutions to the Boltzmann equation (\ref{BE}). We begin by noting that
\begin{eqnarray}\label{Hth}
\frac{dE^{(fB)}}{dt}&=&0\nonumber \\
\frac{dN^{(fB)}}{dt}&=&0\nonumber \\
\frac{dS^{(B)}}{dt}&=&\int d\rr\int d\vv \left[f^*\Xi^{(B)}_{f^*}\right]_{f^*=S^{(B)}_f}>0
\end{eqnarray}
The first equation is a consequence of the Hamiltonian structure of the nondissipative part of the Boltzmann equation and the degeneracy of the dissipation potential $\Xi^{(B)}$. In the second equation, $N^{(fB)}=\int d\rr\int d\vv f(\rr,\vv)$ has the physical interpretation of  the number of moles. The equality holds because $N^{(fB)}$ is the Casimir and because of the degeneracy of the dissipation potential $\Xi^{(B)}$
The third relation constitutes the famous Boltzmann's H-Theorem. The last inequality on its  right hand side  follows directly from: (i) $ [\Xi^{(B)}(f,f^*)]_{X^{(B)}=0}=0$, (ii) $\Xi^{(B)}(f,f^*)$ reaches its minimum at $X^{(B)}=0$, and (iii) $\Xi^{(B)}(f,f^*)$ is a convex function of $f^*$ in a neighborhood of $X^{(B)}=0$. These three properties are directly seen in (\ref{Xi}).
From (\ref{Hth}) we conclude that solutions to (\ref{BE}) approach, as $t\rightarrow \infty$,  solutions to $X^{(B)}=0$. We shall denote the distribution functions that arise as solutions to $X^{(B)}=0$ by the symbol $f_{leq}(\rr,\vv)$, the submanifold that they form by the symbol $\mathcal{M}_{coll}$
\begin{equation}\label{Mcoll}
\mathcal{M}_{coll}=\{f\in M^{(kt)}| X^{(B)}=0\}
\end{equation}
where $M^{(kt)}$ denotes the state space of kinetic theory.

So far, the Casimir $S^{(B)}(f)$ is unspecified. Now we specify it. There are in fact three routes that we can  follow to specify   it. All three routes lead to the same  Boltzmann entropy
(that is one particular Casimir $C(f)$ for which $c(f)=-k_B h^3 f(\ln f-1)$, $k_B$ is the Boltzmann constant and $h$ the Planck constant).

On the first route we ask the question of what is $X^{(B)}$. Since  -  see (\ref{X}) -  $X^{(B)}$ is directly related to  $S^{(B)}(f)$,  we ask in fact the question  of what is $S^{(B)}(f)$)
for which solutions to $X^{(B)}=0$ have the experimentally observed Maxwellian dependence on $\vv$. It is easy to see that the answer is: $S^{(B)}(f)$ is  the Boltzmann entropy.

On the second route we look for the entropy for which the fundamental thermodynamic relation implied by (\ref{BE}) is the fundamental thermodynamic relation representing in equilibrium thermodynamics the ideal gas. This is because the physical system whose time evolution is represented by the Boltzmann kinetic equation (\ref{BE}) is the ideal gas.
We briefly recall (see more in \cite{Gdth}) the way how to derive the fundamental thermodynamic relation implied by a kinetic equation whose solutions have the properties (\ref{Hth}). First, we construct the thermodynamic potential
\begin{equation}\label{Thpot}
\Phi^{(B)}(f,T,\mu)=-S^{(B)}(f)+\frac{1}{T}E^{(fB)}(f)-\frac{\mu}{T}N^{(fB)}(f)
\end{equation}
where $T$ is the equilibrium thermodynamic temperature and $\mu$ the equilibrium thermodynamic chemical potential. Next, we solve the equation $\Phi^{(B)}_{f}(f,T,\mu)=0$. The solution, denoted  $f_{eq}(\rr,\vv,T,\mu)$ and called an equilibrium  distribution,  is approached as $t\rightarrow \infty$. Finally, the fundamental thermodynamic relation $P=P(T,\mu)$ implied by (\ref{BE}) ($P$ is the equilibrium thermodynamic pressure)  is
\begin{equation}\label{PVT} \left[\Phi^{(B)}(f,T,\mu)\right]_{f_{eq}}=\frac{PV}{T} \end{equation}
where $V$ is the volume of the region in which the gas under investigation is confined. It is easy to verify that in order that (\ref{PVT}) is the fundamental thermodynamic relation of the ideal gas, $S^{(B)}(f)$ has to be the Boltzmann entropy.

The third route is the route taken by  Ludwig Boltzmann in his original introduction of the Boltzmann equation. The collision term $\left(\frac{\partial f}{\partial t}\right)_{diss}$ emerges  first  from an analysis of the mechanics of binary collisions. Boltzmann then investigates solutions of his equation by looking  for a potential whose time evolution obeys the inequality in the third equation in (\ref{Hth}). Such search amounts in fact  to casting the collision term, derived initially from mechanics of binary collisions,  into the mass-action-law  form (\ref{collW}).

From the above discussion we see that the  properties (\ref{Hth})   suffice to show agreement with the experimental observation of the approach to the  Maxwellian distribution in velocities but they
do not, by themselves,  suffice to prove the observed approach to the spatially homogeneous distribution.  To prove it, we can ether follow Refs.\cite{Grad},\cite{VillaniB} and investigate further solutions of (\ref{BE}) or follow the spirit of the Boltzmann passage from (\ref{BEfull}) to (\ref{BE}) and make another passage from (\ref{BE}) to a modified Boltzmann equation in which the approach to the spatially homogeneous distribution  becomes manifestly displayed.

Grad \cite{Grad}, Desvillettes  and Villani \cite{VillaniB}  investigate the influence of
the first term on the right hand side of (\ref{BE}) (i.e. the free flow term that leaves  the entropy $S^{(B)}$ unchanged) on solutions of the Boltzmann equation. They have shown that
solutions to Boltzmann equation, \eqref{BE}, approach rapidly a neighborhood of the locally Maxwellian manifold $\mathcal{M}_{coll}$ (see \eqref{Mcoll}). Due to the influence of the free flow term, however, the solutions touch it only at the end of the time evolution, where $f$ does not change in time and is spatially homogeneous, i.e. the distribution $f$ reaches the total Maxwellian distribution, which is the minimum of the thermodynamic potential $\Phi^{(B)}$ (see \eqref{Thpot}).

The second way to investigate the tendency to spatial homogenization in solutions to the Boltzmann equation \eqref{BE}  is  motivated by another experimental observation made on the Boltzmann gas. According to this   observation  the time evolution of the Boltzmann gas becomes, as $t\rightarrow \infty$,  well described by fluid mechanics that is a mesoscopic theory in which  more microscopic details are ignored than they are ignored in the kinetic theory.
The proof consists of showing first that solutions to (\ref{BE})
become, as $t\rightarrow\infty$, well approximated by  solutions to the governing equations of fluid mechanics and then, in the second part, showing that  solutions to the governing equations of fluid mechanics approach, again as $t\rightarrow\infty$,  spatially homogeneous solutions. Regarding  the first part of the proof, there are several methods  (the most well known among them is the Chapman-Enskog method) that can be used to reduce kinetic theory to fluid mechanics. All the methods display well the underlying physics  but their  mathematical rigor is inferior to the mathematical rigor with which the results  in \cite{VillaniB} are derived. Another way to transform the Boltzmann equation (\ref{BE}) into an equation manifestly displaying approach to spatially homogeneous solution can be an adaptation to the Boltzmann equation of the modification introduced in the next section for the Vlasov equation. We shall return to this possibility  later in this paper.

In order to find still more information about solutions to the Boltzmann equation, we can restrict ourselves to particular stages in the time evolution and correspondingly simplify the Boltzmann equation. For example, in the final stages of the approach to the thermodynamic equilibrium states $f_{eq}(\rr,\vv)$ the thermodynamic force $X^{(B)}$  (see (\ref{X})) is small and we can thus approximate the dissipation potential $\Xi^{(B)}$ (see (\ref{Xi})) by $ \Xi^{(B)}(f,f^*)=\int d\rr\int d\vv\int d\vv'W^{(B)}(X^{(B)})^2$. If moreover the changes of the momenta in the binary collisions are small, we can approximate $X^{(B)}$ by $X^{(B)}=\frac{\partial f^*}{\partial \vv}$. With these approximations,
$\left(\frac{\partial f}{\partial t}\right)_{diss}$ takes  the Fokker-Planck form
\begin{equation}\label{FP}
\left(\frac{\partial f}{\partial t}\right)_{diss}=\frac{\partial}{\partial \vv}\left(\Lambda^{(B)}f\frac{\partial f^*}{\partial \vv}\right)
\end{equation}
where $\Lambda^{(B)}$ is an operator that is  degenerate (in order that  the energy is preserved during the time evolution) and positive definite if acting  outside of its nullspace, and $f^*=\Phi_{f}$.

An additional physical insight into the Boltzmann regularization can be gained by regarding the kinetic equations (\ref{BEfull}) and (\ref{BE}) as Liouville equations corresponding to equations governing the time evolution of one quasi-particle   and then interpreting the regularization as a modification of the quasi-particle dynamics. We now proceed to identify the quasi-particle dynamics.

We begin by noting that  entropy conservation in the nondissipative time evolution allows us to write
the kinetic  equation (\ref{BEfull}) in the form
\begin{equation}\label{BEfullstar}
\frac{\partial f}{\partial t}=-\frac{\partial}{\partial \rr}\left( Tf\frac{\partial f^*}{\partial\vv}\right)+\frac{\partial}{\partial \vv}\left( Tf\frac{\partial f^*}{\partial\rr}\right)
\end{equation}
and the Boltzmann equation (\ref{BE})  (if we use the Fokker-Planck form (\ref{FP}) of the Boltzmann collision term) in the form
\begin{equation}\label{BEstar}
\frac{\partial f}{\partial t}=-\frac{\partial}{\partial \rr}\left(Tf\frac{\partial f^*}{\partial\vv}\right)+\frac{\partial}{\partial \vv}\left( Tf\frac{\partial f^*}{\partial\rr}\right)+\frac{\partial}{\partial \vv}\left( f\Lambda^{(B)}  \frac{\partial f^*}{\partial \vv}\right)
\end{equation}
Note that this dissipation is very similar to the Landau collision integral \cite{LL10}, \S 41. In the case of the general Boltzmann collision term (\ref{collW}) we still can write the Boltzmann kinetic equation in the form (\ref{BEstar}) (see \cite{Grcontmath})  if we use the identity
\begin{equation}\label{Kid}
\int dy'[\varphi(y',y)-\varphi(y,y')]=-\frac{\partial}{\partial y_j}\int dy'\int d\eta
y'\varphi(y(\eta),y'(\eta))
\end{equation}
where $\varphi$ is a sufficiently regular mapping
$\varphi:{\mathbb{R}}^N\rightarrow{\mathbb{R}}^N$,
$y(\eta)=y-\eta y'; \, y'(\eta)=y+(1-\eta)y'$ used in particular in \cite{Kirkwood}

Next, we regard the above two kinetic equations as
Liouville equations corresponding to
\begin{eqnarray}\label{pBEfull}
\dot{\rr}&=&T\frac{\partial f^*}{\partial \vv}=\frac{\vv}{m}\nonumber \\
\dot{\vv}&=&-T\frac{\partial f^*}{\partial \rr}
\end{eqnarray}
in the case of (\ref{BEfullstar}) and
\begin{eqnarray}\label{pBE}
\dot{\rr}&=&T\frac{\partial f^*}{\partial \vv}=\frac{\vv}{m}\nonumber \\
\dot{\vv}&=& \Lambda^{(B)}\frac{\partial f^*}{\partial \vv} -T \frac{\partial f^*}{\partial \rr}
\end{eqnarray}
in the case of the Boltzmann equation (\ref{BEstar}). Both the quasi-particle-dynamics equations (\ref{pBEfull}) and (\ref{pBE}) involve the distribution function $f$ and have to be therefore considered always together with their corresponding Liouville equations (\ref{BEfullstar}), in the case of (\ref{pBEfull}),  and (\ref{BEstar}), in the case of (\ref{pBE}).

The regularization of the kinetic equation (\ref{BEfull}) into the Boltzmann equation (\ref{BE}) we can now see as replacing the force $T\frac{\partial f^*}{\partial \rr}$ in the quasi-particle dynamics (\ref{pBEfull}) with the friction force $\Lambda^{(B)}\frac{\partial f^*}{\partial \vv}$. Note that this friction conserves the energy (see e.g. \cite{LL10}, \S 41), it is a friction in the sense that it causes  an increase of  the entropy.

\section{Vlasov kinetic theory}\label{3}

Now we turn to the main subject of this paper. We consider the Vlasov gas and the Vlasov kinetic equation governing its time evolution. The Vlasov equation  has still the form (\ref{KE}) with the Poisson bivector  given by (\ref{L}) and the energy
\begin{eqnarray}\label{enpotV}
E^{(fV)}&=&\int d\rr\int d\vv f(\rr,\vv)\frac{\vv^2}{2m}+E^{(fV)}_{(pot)}(f)\nonumber \\
&=&\int d\rr\int d\vv f(\rr,\vv)\frac{\vv^2}{2m}\nonumber \\
&&+\frac{1}{2}\int d\rr_1\int d\vv_1 \int d\rr_2\int d\vv_2 f(\rr_1,\vv_1)f(\rr_2,\vv_2)V_{pot}^{(fV)}(|\rr_2-\rr_1|)\nonumber \\
\end{eqnarray}
where $V_{pot}^{(fV)}(|\rr_2-\rr_1|)$ is the two-particle long range attractive potential.  Written explicitly, the Vlasov kinetic equation has the form
\begin{eqnarray}\label{Vlasov}
\frac{\partial f(\rr,\vv)}{\partial t}&=& L^{(f)}E^{(fV)}_f\nonumber \\
&=&-\frac{\vv}{m}\frac{\partial f(\rr,\vv)}{\partial \rr}+\frac{\partial}{\partial\vv}\left(f(\rr,\vv)\frac{\partial \int d\rr_1\int d\vv_1 V^{(fV)}(|\rr-\rr_1|)
f(\rr_1,\vv_1)}{\partial\rr}\right)\nonumber \\
\end{eqnarray}
We see that the equations governing the time evolution of the Boltzmann gas (Eq.\eqref{BEfull}) and the equation governing the time evolution of the Vlasov gas (Eq.\eqref{Vlasov}) are the same except that the potential energy in the case of the Boltzmann gas is short range and repulsive and in the case of the Vlasov gas long range and attractive.

As for the solutions to (\ref{Vlasov}),  we  note, as we did in the case of \eqref{BEfull},   that (\ref{Vlasov}) is a particular realization of the general kinetic equation (\ref{KE}) and thus the four properties of solutions derived in Section \ref{1}
\begin{eqnarray}\label{BEfthV}
&&\frac{dE^{(fV)}}{dt}=0\nonumber \\
&&\frac{dC^{(V)}}{dt}=0\nonumber \\
&&\frac{dN^{(V)}}{dt}=0\nonumber \\
&& time\,\,reversibility
\end{eqnarray}
where  $C^{(V)}$ is a Casimir of the bracket (\ref{L}), and $N^{(V)}(f)=\int d\rr\int d\vv f(\rr,\vv)$ is another Casimir, that has the physical interpretation of the number of moles, hold also for (\ref{Vlasov}). Regarding more specific properties, we shall concentrate on the experimentally observed \cite{expLandau} tendency to spatial homogenization known as Landau damping. It has been rigorously proven in \cite{VillaniV} that solutions to (\ref{Vlasov}), both with and without the second term on its right hand side, converge weakly, as $t\rightarrow\infty$,  to spatially homogeneous distribution.
Inspired by the Boltzmann regularization of (\ref{BEfull}),
our objective in this paper is to modify  (\ref{Vlasov}) into a new kinetic equation in which the tendency to the spatial homogenization is manifestly displayed.

\subsection{Extension }\label{Vext}

As in the Boltzmann regularization, our first task is to attempt to recognize the physics that is behind the property of  solution of the governing equation that is in the focus of our interest. In the case of the Vlasov equation it is the Landau damping (i.e. the time irreversible tendency towards the spatial homogenization).
There does not seem to be a general agreement about  the physics behind the Landau damping   (see e.g. Section 1.2   in \cite{VillaniV}) but there  does seem  to be  an agreement that, due to the phase mixing, the particles experience a complex motion that can be regarded  as   micro-turbulence (i.e. the turbulence seen on the level of kinetic theory with the one particle distribution function serving as the state variable) \cite{Microturb1,Microturb2}.

In the rest of this paper we   extend the Vlasov kinetic theory. By an extension we mean first of all an enlargement of the state space. How do we choose the extra state variables? In principle, we choose them as the state variables through which we can best and in the most succinct  way express the new features that are missed in the original non extended theory. In practice, the extra state variables are often found in the vector field of the original dynamical theory (as it is done in the Grad type extensions of the fluid mechanics in \cite{MullRugg}, \cite{Jouetal}, \cite{RuggA} by adopting hydrodynamic fluxes as the extra state variables) or in the investigation of the more detailed (more microscopic) nature of the macroscopic system under investigation (as it is done in fluid mechanics of complex fluids for example in \cite{Kirkwood}, \cite{Bird}). The former type of extension is, from the physical point of view, an extension by introducing an extra inertia, the latter type of the extension is, from the physical point of view, an involvement of the internal structure whose time evolution cannot be decoupled from the time evolution of the ineterst in the original dynamical theory.

Having chosen the extra state variables, we face the problem of establishing equations governing their time evolution. In the Grad type extensions this is typically done by suggesting closures of the infinite Grad hierarchy. In the internal-structure type extensions the time evolution equations are typically found, first, by investigating the time evolution of the internal structure (e.g. the time evolution of a dumbbell suspended in a fluid - see e.g. \cite{Bird}), and second, by investigation the coupling between the time evolution of the original and the time evolution of the extra state variables (expressed, for example in the context of fluid mechanics of complex fluids in the dependence of the stress tensor on the extra state variables). A complementary and a more systematic way of finding the time evolution equations of an extended theory is to require that both the original and the extended dynamical theories share the mathematical structure guaranteeing agreement of their solutions with certain basic experimental observations.

In our extension of the Vlasov kinetic theory we shall follow the internal-structure type extension and we shall require that the time evolution in both the original and the extended theory is Hamiltonian.
The extended kinetic theory that we shall formulate below involves an extra state variable that is chosen to be the one particle distribution function characterizing the internal structure induced by the micro-turbulence. In the nutshell, we replace    the distribution function $f(\rr,\vv)$ with a pair of distribution functions
\begin{equation}\label{Reynolds}
(\phi(\rr,\vv),\psi(\rr,\vv))
\end{equation}
We interpret $\phi(\rr,\vv)$ as the average (or regularized) distribution function and $\psi(\rr,\vv)$ as a distribution function characterizing the internal structure. The energy (\ref{enpotV}) acquires  in the extended theory a new term expressing the extra kinetic energy involved in the small scale micro-turbulent motion. The extended Vlasov kinetic equation (i.e. the equation governing the time evolution of (\ref{Reynolds}) is Hamiltonian and its solutions satisfy the properties (\ref{BEfthV}) (appropriately adapted to the extended theory).   The regularization (discussed in Section \ref{Vreg}) of the extended Vlasov kinetic equation is then made by adding to it
a Fokker-Planck like dissipation of the micro-turbulent structure. In the course of the time evolution, this dissipation induces  then the Landau damping of  $\phi(\rr,\vv)$.

We now proceed to introduce the equation governing the time evolution of (\ref{Reynolds}), i.e. the equation extending the Vlasov equation (\ref{Vlasov}). The physical requirement on which we base our construction is the preservation of the Hamiltonian structure and of the properties (\ref{BEfthV}) of solutions of the original Vlasov equation \eqref{Vlasov}.  We therefore begin by extending the Poisson bracket (\ref{L}) to the Poisson bracket expressing kinematics of (\ref{Reynolds}). In order to see well both the physics and the mathematics that is behind the extension, we shall make the extension by following  two routes. On the first route (Section \ref{1B}),  the extension  is motivated by the geometrical content of the bracket \eqref{L}, on the second route (Section \ref{Reyext}),  by the Reynolds approach to the turbulence in fluid mechanics.

\subsubsection{Geometrical approach}\label{1B}

 From the physical point of view, we  assume that the kinematics is still determined by the kinematics of $\phi(\rr,\vv)$ expressed in the Poisson bracket (\ref{L}) (in which $f$ is replaced by $\phi$) and that the newly adopted  distribution function $\psi(\rr,\vv)$ is passively advected. From the mathematical point of view,  we shall make therefore the extension with the help of the concept of  semidirect product (see e.g. \cite{MRat}, \cite{Ogul}).

Poisson bracket \eqref{L} can be regarded as the Lie-Poisson bracket on the dual of the Lie algebra of Hamiltonian vector fields on the contangent bundle $\mathcal{T}^*Q$, $Q\subset\mathbb{R}^3$ of one-particle classical mechanics, see \cite{MRat}. The vector fields
\begin{equation}\label{vfM}
\XX_h = \frac{\partial h}{\partial \vv}\cdot\frac{\partial}{\partial \rr}
-\frac{\partial h}{\partial \rr}\cdot\frac{\partial}{\partial \vv},
\end{equation}
where $h$ is the one particle Hamiltonian, generate the Lie group $G$ of canonical transformations on  $\mathcal{T}^*Q$. Hence, a natural extension of the dynamics generated by \eqref{L} is a geometric extension (semidirect product) of the underlying Lie group.

The vector field \eqref{vfM}, which depends on the choice of the Hamiltonian $h$, generates motion of a particle within the cotangent bundle and that motion can be regarded as a Lie group. The vector field form an algebra called Lie algebra of $G$ and denoted by $\mathcal{G}$. Bracket \eqref{L} can be seen as the Lie-Poisson bracket governing evolution in the dual of $\mathcal{G}$, where fields of differential forms live. The distribution function $\phi$ can be then introduced as the phase-space divergence of the differential forms, and bracket \eqref{L} then generates evolution equations for $\phi$.

The Lie group $G$ can be extended by adding an another group passively advected by $G$. That is the semidirect product, and according to Eq. 3.1 in \cite{MW} the Poisson bracket constructed from the semidirect product $G\rtimes V$, which gives evolution on the Lie algebra dual of the semidirect product, is
\begin{eqnarray}\label{eq.EKT.bracket}
\{A,B\}^{(ext)} &=&\langle \phi, [A_{\phi},B_{\phi}]\rangle + \langle \psi, -\XX_{A_{\phi}}(B_\psi)\rangle- \langle \psi, -\XX_{B_{\phi}}(A_\psi)\rangle
\end{eqnarray}
The new density  $\psi(\rr,\vv)$ is the density expressing the added (passively advected) group. Let us refer to this bracket as to the extended kinetic theory Poisson bracket. The bracket $[\bullet,\bullet]$ stands for the Poisson bracket of classical mechanics, which governs evolution on $\mathcal{T}^Q$,
\begin{equation}
[a,b]=\frac{\partial a}{\partial \rr}\cdot\frac{\partial b}{\partial \vv}
-\frac{\partial b}{\partial \rr}\cdot\frac{\partial a}{\partial \vv} \qquad\forall a,b\in V.
\end{equation}
Bracket \eqref{eq.EKT.bracket} can be then rewritten more explicitly as

\begin{eqnarray}\label{brext}
\{A,B\}^{(ext)} &=&\int\dd\rr\int\dd\vv \phi \left(\frac{\partial A_{\phi}}{\partial \rr}\frac{\partial B_{\phi}}{\partial \vv}
-\frac{\partial B_{\phi}}{\partial \rr}\frac{\partial A_{\phi}}{\partial \vv}\right)\nonumber\\
&&+ \int\dd\rr\int\dd\vv \psi \left(\frac{\partial A_{\phi}}{\partial \rr}\cdot \frac{\partial B_\psi}{\partial \vv}
-\frac{\partial B_{\phi}}{\partial \rr}\cdot \frac{\partial A_\psi}{\partial \vv}\right)\nonumber\\
&&+ \int\dd\rr\int\dd\vv \psi \left(\frac{\partial A_\psi}{\partial \rr}\cdot \frac{\partial B_{\phi}}{\partial \vv}
- \frac{\partial B_\psi}{\partial \rr}\cdot \frac{\partial A_{\phi}}{\partial \vv}\right).
\end{eqnarray}

The  evolution equations of $\phi$ and $\psi$ can be  obtained by casting $\{A,T \Phi\}^{(ext)}$ into the form
\begin{equation}
\int\dd\rr\int\dd\vv\left( A_{\phi} \frac{\partial \phi}{\partial t} + A_\psi \frac{\partial \psi}{\partial t}\right),
\end{equation}
which leads to
\begin{subequations}\label{eq.EKT.evo}
\begin{eqnarray}
\frac{\partial \phi}{\partial t} &=& -\frac{\partial}{\partial \rr}\cdot\left(\phi T\frac{\partial \phi^*}{\partial \vv}\right)
+\frac{\partial}{\partial \vv}\cdot\left(\phi T\frac{\partial \phi^*}{\partial \rr}\right)\nonumber\\
&&-\frac{\partial}{\partial \rr}\left(\psi T\frac{\partial \psi^*}{\partial \vv}\right)
+\frac{\partial}{\partial \vv}\left(\psi T\frac{\partial \psi^*}{\partial \rr}\right)\\
\frac{\partial \psi}{\partial t} &=&-\frac{\partial}{\partial \rr}\left(\psi T\frac{\partial \phi^*}{\partial \vv}\right)
+\frac{\partial}{\partial \vv}\left(\psi T\frac{\partial \phi^*}{\partial \rr}\right)
\end{eqnarray}
\end{subequations}
where $\phi^*=\Phi_{\phi}$ and $\psi^*=\Phi_{\psi}$. These are the governing equations of the extended Vlasov theory.  They are Hamiltonian as the original Vlasov equations and their solutions have the properties
\begin{eqnarray}\label{HthVV}
&&\frac{dE^{(ext)}}{dt}=0\nonumber \\
&&\frac{dN^{(ext)}}{dt}=0 \nonumber \\
&&\frac{dC^{(ext)}}{dt}=0 \nonumber \\
&&time\,\,reversibility
\end{eqnarray}
where $C^{(ext)}$ is a Casimir of the Poisson bracket (\ref{brext}) and $N^{(ext)}$ is the Casimir having the physical interpretation of the number of moles, that are the same as (\ref{BEfthV}).

Regarding the energy $E^{(ext)}(\phi,\psi)$, the physical interpretation of (\ref{Reynolds}) leads us to
\begin{eqnarray}\label{enturb}
E^{(ext)}(\phi,\psi)&=&\int d\rr\int d\vv \phi(\rr,\vv) \frac{\vv^2}{2m}\nonumber \\
&&+\frac{1}{2}\int d\rr_1\int d\vv_1 \int d\rr_2\int d\vv_2 \phi(\rr_1,\vv_1)\phi(\rr_2,\vv_2)V_{pot}^{(fV)}(|\rr_2-\rr_1|)\nonumber \\
&&+E^{(turb)}(\phi, \psi)
\end{eqnarray}
The new contribution $E^{(turb)}(\phi, \psi)$ to the total energy $E^{(ext)}(\phi,\psi)$ is the energy associated with the micro-turbulence. It is therefore mainly an extra  kinetic energy. The total kinetic energy is thus a sum of the overall kinetic energy $\int d\rr\int d\vv \phi(\rr,\vv) \frac{\vv^2}{2m}$, that depends only on the overall distribution function $\phi$,  and the extra kinetic energy $E^{(turb)}(\phi,\psi)$ of the fine scale micro-turbulent motion. For the qualitative analysis of solutions to (\ref{eq.EKT.evo}), presented  in Section \ref{Vreg},  we do not need a specific form of  $E^{(turb)}(\phi,\psi)$. We assume that the contribution of the  micro-turbulence to the potential energy of the Vlasov gas is negligible.
The potential energy given by the second term in (\ref{enturb})  is thus assumed to represent the complete  potential energy also in the extended Vlasov theory.

\subsubsection{Reynolds approach adapted to kinetic theory}\label{Reyext}

Our objective is to adapt the classical Reynolds approach to the turbulence (developed originally by Reynolds \cite{Reynolds} inside  fluid mechanics) to the micro-turbulence of the Vlasov gas (formulated now inside kinetic theory). We anticipate that in  the  Reynolds-type reformulation of the Vlasov equation we will be  able to see the Landau damping similarly as the extra Reynolds stress (arising due to the turbulence) is seen in the  Reynolds analysis.

We begin by recalling the essential steps that are made in the original Reynolds analysis. The starting point is the Navier-Stokes equation governing the time evolution of the velocity field that plays the role of the state variable. In the first step,  this equation is extended to a more microscopic level of description (i.e. to a  level on which more details are taken into account). This is done by promoting the velocity field to the status of the  random variable and extending the Navier-Stokes equation to the corresponding to it Langevin equation. Alternatively, the extension of the velocity field can be done by representing  the random velocity field as  the average field together with  an infinite  hierarchy of moments that are physically interpreted as fields characterizing the internal structure. The Langevin equation governing the time evolution of the random velocity field then
provides the equations governing the time evolution of this extended set of state variables.
The subsequent  passage from the  infinite to a finite number of moments requires a closure of the infinite hierarchy of the time evolution equations.

We take the above description of the Reynolds analysis as a motivation rather than a recipe.
The starting point in our Reynolds-type investigation of the micro-turbulence is the Vlasov equation (\ref{Vlasov}). Our first step is not the promotion of the one particle distribution function to the status of random variable   but rather its  replacement by a pair of distribution functions \eqref{Reynolds}.  Both distribution functions $\phi(\rr,\vv)$ and $\psi(\rr,\vv)$ are regarded as being  independent. The equations governing their time evolution are then  constructed by requiring that the Hamiltonian structure of the original Vlasov equation (\ref{Vlasov}) is preserved in the extended time evolution equations.

If we put the Reynolds analysis  into  the setting of the Hamiltonian dynamics and  recall that the state variable in the Hamiltonian formulation of the Euler equation is the momentum field (see e.g. \cite{MW},  \cite{HamNS}) and that the velocity field is its conjugate,
we see that the essential step  in the Reynolds analysis, namely the separation into an average and fluctuations, is not made for the state variable but for its conjugate. This means that in the setting of kinetic theory
we shall replace   the distribution function $f(\rr,\vv)$ with a pair of distribution functions \eqref{Reynolds}
in such a way that
\begin{equation}\label{12star}
f^*=\phi^*+\psi^*
\end{equation}

In order to arrive at the Poisson bracket expressing the kinematics of \eqref{Reynolds},  we proceed as follows. First, we introduce two uncoupled and identical Vlasov equations, one for the distribution function $f$ and the other for the distribution function $g$. The Poisson bracket expressing the kinematics of $(f,g)$ is simply
\begin{equation}\label{br12}
\{A,B\}^{(fg)}=\{A,B\}^{(f)}+\{A,B\}^{(g)}
\end{equation}
where $\{A,B\}^{(f)}$ is the bracket (\ref{L}) and $\{A,B\}^{(g)}$ is the same bracket but with $f$ replaced by $g$.
Now, we look for a one-to-one transformation $(f,g)\leftrightarrows (\phi,\psi)$ for which $f^*=\phi^*+\psi^*$ (which is the relation (\ref{12star})) and $g^*=\phi^*$. We see easily that such transformation is
\begin{eqnarray}\label{transf}
\phi &=&f+g\nonumber \\
\psi &=&f
\end{eqnarray}
If we now apply this transformation to (\ref{br12}) we obtain
\begin{equation}\label{brextR}
\{A,B\}^{(\phi\psi)}=\{A,B\}^{(\phi)}+\{A,B\}^{(\psi )}+\widehat{\{A,B\}}^{(\phi,\psi )}
\end{equation}
where $\{A,B\}^{(\phi)}$  is the Poisson bracket (\ref{L}) with $f$ replaced by $\phi$,   $\{A,B\}^{(\psi )}$ is the Poisson bracket (\ref{L}) with $f$ replaced by $\psi$, and
\begin{eqnarray}\label{br11}
\widehat{\{A,B\}}^{(\phi,\psi )}&=&\int d\rr\int d\vv  \psi\left[\frac{\partial A_{\psi}}{\partial \rr}\frac{\partial B_{\phi}}{\partial \vv}-\frac{\partial B_{\psi}}{\partial \rr}\frac{\partial A_{\phi}}{\partial \vv}\right]\nonumber \\
&&+ \int d\rr\int d\vv \psi \left[\frac{\partial A_{\phi}}{\partial \rr}\frac{\partial B_{\psi}}{\partial \vv}-\frac{\partial B_{\phi}}{\partial \rr}\frac{\partial A_{\psi}}{\partial \vv}\right]
\end{eqnarray}
Since the transformation (\ref{transf}) is one-to-one and (\ref{br12}) is a Poisson bracket then also the bracket (\ref{brext}) is a Poisson bracket (in particular, we are certain that the Jacobi identity holds for \eqref{brext}). If we now compare the  bracket \eqref{brextR} with the bracket \eqref{brext}, we see that  $\{A,B\}^{(\phi\psi)} =\{A,B\}^{(ext)}+\{A,B\}^{(\psi)}$. We have proven that both $\{A,B\}^{(\phi\psi)}$ and $\{A,B\}^{(ext)}$ are Poisson brackets. The equations governing  the time evolution of \eqref{Reynolds} implied by  $\{A,B\}^{(\phi\psi)}$ and by $\{A,B\}^{(ext)}$ are the same except  that the equation governing the time evolution of $\psi(\rr,\vv)$ that is implied by $\{A,B\}^{(\phi\psi)}$  involves two extra terms: $\frac{\partial}{\partial \rr}\left(T\psi\frac{\partial \psi^*}{\partial\vv}\right) +\frac{\partial}{\partial \vv}\left(T\psi\frac{\partial  \psi^*}{\partial\rr}\right)$.

\subsection{Regularization}\label{Vreg}

In order to get more detailed properties of solutions to the extended Vlasov equation (\ref{eq.EKT.evo}) (more detailed than (\ref{HthVV})),  we can either  follow Desvillettes and Villani  \cite{VillaniV} and adopt their rigorous mathematical analysis of the Vlasov equation (\ref{Vlasov}) to the extended Vlasov equation (\ref{eq.EKT.evo}) or we can follow Reynolds and, first,  regularize (\ref{eq.EKT.evo}),  and then  investigate solutions of the regularized   equation.  The regularization   of  (\ref{exteqR}) consists of supplying  (\ref{eq.EKT.evo}) with an explicit dissipation term expressing the physics of the micro-turbulence. We shall follow the latter route.

We  assume that the thermodynamic force $X^{(V)}$  (compare with \eqref{X}) that drives  the decay of the micro-turbulence is effective velocity $\frac{\partial \psi^*}{\partial \vv}$. We limit ourselves to the quadratic dissipation potential
\begin{equation}\label{XiV}
\Xi^{(V)}(\phi,\psi,\psi^*)=\int d\rr\int d\vv \psi\Lambda^{(V)}(X^{(V)})^2
\end{equation}
where $\Lambda^{(V)}$ is a symmetric, degenerate (in order the total energy remains conserved, i.e. $\Lambda^{(V)}\frac{\partial E^{(ext)}_{\psi}}{\partial \vv}=0$) and positive definite if acting outside its nullspace.

The regularized extended Vlasov kinetic equation  becomes
\begin{subequations}\label{exteqR}
\begin{eqnarray}
\frac{\partial \phi}{\partial t}&=&-\frac{\partial}{\partial \rr}\left(T\phi\frac{\partial \phi^*}{\partial\vv}\right) +\frac{\partial}{\partial \vv}\left(T\phi\frac{\partial  \phi^*}{\partial\rr}\right)\nonumber \\
&&-\frac{\partial}{\partial \rr}\left(T\psi\frac{\partial \psi^*}{\partial\vv}\right) +\frac{\partial}{\partial \vv}\left(T\psi\frac{\partial  \psi^*}{\partial\rr}\right) \\
\frac{\partial \psi}{\partial t}&=&-\frac{\partial}{\partial \rr}\left(T\psi\frac{\partial \phi^*}{\partial\vv}\right) +\frac{\partial}{\partial \vv}\left(T\psi\frac{\partial  \phi^*}{\partial\rr}\right)\nonumber \\
&&+\frac{\partial}{\partial\vv}\left(\psi \Lambda^{(V)}\frac{\partial\psi^*}{\partial \vv}\right)
\end{eqnarray}
\end{subequations}
The dissipative term in the last line is $-\Xi^{(V)}_{\psi}=\frac{\partial}{\partial\vv}\left(\psi \Lambda^{(V)}\frac{\partial\psi^*}{\partial \vv}\right)$.

Solutions to (\ref{exteqR}) have the following properties:
\begin{eqnarray}\label{HthVVdiss}
\frac{dE^{(ext)}}{dt}&=& 0\nonumber \\
\frac{dN^{(ext)}}{dt}&=& 0 \nonumber \\
\frac{dS^{(V)}}{dt}&=& \int d\rr\int d\vv \left[\psi^*\Xi^{(V)}_{\psi^*}\right]_{\psi^*=S^{(V)}_{\psi}}\nonumber \\
&&=\int d\rr\int d\vv\frac{\partial S^{(V)}_{\psi}}{\partial \vv}\Lambda^{(V)}
\frac{\partial S^{(V)}_{\psi}}{\partial \vv}>0
\end{eqnarray}

To continue the  investigation of solutions to (\ref{exteqR}), we regard it as a pair of  Liouville equations  corresponding to two quasi-particle time evolution equations (compare with  (\ref{pBE})). The equation governing the time evolution of the "average" quasi-particle are
\begin{eqnarray}\label{pVphi}
\dot{\rr}&=&T\frac{\partial \phi^*}{\partial \vv}+T\frac{\psi}{\phi}\frac{\partial\psi^*}{\partial \vv}\nonumber \\
\dot{\vv}&=& -T\frac{\partial \phi^*}{\partial\rr}-T\frac{\psi}{\phi}\frac{\partial\psi^*}{\partial \rr}
\end{eqnarray}
and the equations governing the time evolution of the  "turbulence" quasi-particle are
\begin{eqnarray}\label{pVpsi}
\dot{\rr}&=&T\frac{\partial\phi^*}{\partial \vv}\nonumber \\
\dot{\vv}&=& -T\frac{\partial\phi^*}{\partial \rr}- \Lambda^{(V)}\frac{\partial \psi^*}{\partial\vv}
\end{eqnarray}

Now we use our  physical insight into the micro-turbulence
and estimate the relative importance of the terms appearing on the right hand side of (\ref{pVphi}) and (\ref{pVpsi}).

From the assumptions that we made about the energy $E^{(ext)}$ (see (\ref{enturb})), we can assume that
the extra force $T\frac{\partial \psi^*}{\partial \rr}$ brought about by the micro-turbulence is  small relative to the force $T\frac{\partial \phi^*}{\partial \rr}$
brought about by the interaction potential $V_{pot}^{\phi V}$. On the contrary, the extra velocity of the "average" quasi-particle that is proportional to $\frac{\partial \psi^*}{\partial \vv}$ and that arises due to the micro-turbulence, is not, in general,  small relative to the velocity $T\frac{\partial\phi^*}{\partial\vv}$. Moreover, we can assume that the  "turbulent" quasi-particle has a small mass and thus we we can limit ourselves to the  motion with zero acceleration. This means (if we omit the force $T\frac{\partial \phi^*}{\partial \rr}$) that
\begin{equation}\label{41}
T\frac{\partial\phi^*}{\partial \rr}+ \Lambda^{(V)}\frac{\partial \psi^*}{\partial\vv}=0
\end{equation}
If we now insert this equation to the first equation in (\ref{pVphi}), we obtain
\begin{equation}\label{42}
\dot{\rr}=T\frac{\partial \phi^*}{\partial \vv}-D^{(L)}\frac{\partial\phi^*}{\partial \rr}
\end{equation}
where
\begin{equation}\label{43}
D^{(V)}=T^2\frac{\psi}{\phi}(\Lambda^{(V)})^{-1}
\end{equation}
is the diffusion coefficient of the Landau damping. With (\ref{42}) and with omitting the extra force $T\frac{\partial \phi^*}{\partial \rr}$ in the second equation in (\ref{pVphi}), the equation governing the time evolution of the "average" distribution function $\phi(\rr,\vv)$ becomes
\begin{eqnarray}\label{44}
\frac{\partial \phi}{\partial t}&=&-\frac{\partial}{\partial \rr}\left(T\phi\frac{\partial \phi^*}{\partial\vv}\right) +\frac{\partial}{\partial \vv}\left(T\phi\frac{\partial  \phi^*}{\partial\rr}\right)\nonumber \\
&&+\frac{\partial}{\partial\rr}\left(D^{(V)}\phi\frac{\partial\phi^*}{\partial\rr}\right)
\end{eqnarray}
This is the Vlasov equation (\ref{Vlasov}) equipped with the diffusion term (the last term on the right hand side of (\ref{44})) that manifestly displays the Landau damping.

\section{Spatial homogenization in the extended fluid mechanics }\label{4}

As we noted in Introduction, it is  already  surprising that the one particle kinetic theory has been found to be  microscopic enough to provide an appropriate  setting for describing  dynamics of the Vlasov gas. We therefore do not expect  that such description  can be made in the even less microscopic setting
 of the classical fluid mechanics.  Nevertheless, we can still  ask the question as to whether there exist  extensions of the classical fluid mechanics  which predict spatial homogenization (in the sense that we have seen above in the setting of kinetic theory) and  could  thus become  candidates for the setting suitable for the Vlasov gas. In this section we define what we mean by Landau damping in fluid mechanics (in the last paragraph before Section \ref{VHext}) and introduce two  extensions predicting it.

We begin by recalling the Hamiltonian formulation  of the classical fluid mechanics.
The fields playing the role of the  state variables are
\begin{equation}\label{fmsv}
(\rho(\rr),\uu(\rr),s(\rr))
\end{equation}
denoting the mass, momentum, and entropy density respectively. Their  kinematics is  expressed mathematically in the Poisson bracket (see e.g. \cite{MW})
\begin{eqnarray}\label{5br}
\{A,B\}^{(hyd)}&=&\int d\rr \left[\rho\left(\frac{\partial}{\partial r_{\alpha}}(A_{\rho})B_{u_{\alpha}}-\frac{\partial}{\partial r_{\alpha}}(B_{\rho})A_{u_{\alpha}}\right)\right.\nonumber \\
&&\left.+s\left(\frac{\partial}{\partial r_{\alpha}}(A_{s})B_{u_{\alpha}}-\frac{\partial}{\partial r_{\alpha}}(B_{s})A_{u_{\alpha}}\right)\right.\nonumber \\
&&\left.+u_i\left(\frac{\partial}{\partial r_{\alpha}}(A_{u_i})B_{u_{\alpha}}-\frac{\partial}{\partial r_{\alpha}}(B_{u_i})A_{u_{\alpha}}\right)\right]
\end{eqnarray}
We use hereafter the summation convention; $i=1,2,3; \alpha=1,2,3$. The time evolution equations $\dot{A}=\{A,E\}^{(hyd)}; \forall A$, where $E(\rho,\uu,s)$ is the energy,   are the familiar equations of the Euler fluid mechanics
\begin{eqnarray}\label{5eqs}
\frac{\partial \rho}{\partial t}&=& - \frac{\partial J^{(\rho)}_{\alpha}}{\partial r_{\alpha}}\nonumber \\
\frac{\partial u_i}{\partial t}&=& - \frac{\partial J^{(u)}_{i \alpha }}{\partial r_{\alpha}}          \nonumber \\
\frac{\partial s}{\partial t}&=&- \frac{\partial J^{(s)}_{\alpha}}{\partial r_{\alpha}}\nonumber \\
\frac{\partial e}{\partial t}&=&- \frac{\partial J^{(e)}_{\alpha}}{\partial r_{\alpha}}\nonumber \\
\end{eqnarray}
with
\begin{eqnarray}\label{5flux}
\JJ^{(\rho)}&=&\rho E_{\uu}\nonumber \\
\JJ^{(u)}&=&\ssigma+p\ddelta\nonumber \\
\JJ^{(s)}&=&s E_{\uu}\nonumber \\
\JJ^{(e)}&=&(e+p)E_{\uu}
\end{eqnarray}
where
\begin{equation}\label{sig5}
\ssigma=\uu E_{\uu}
\end{equation}
and
\begin{equation}\label{pp5}
p=-e+\rho E_{\rho}+sE_s+<\uu, E_{\uu}>
\end{equation}
The last equation in (\ref{5eqs}) is a consequence  of the three equations above it (it is a companion local conservation law) but we have added it for the  completeness.

Next, we investigate solutions to \eqref{5eqs}. First, we note that the properties
\eqref{BEfth} (also  \eqref{BEfthV}, or \eqref{HthVV}) hold also for solutions to \eqref{5eqs}. The thermodynamic potential  $\Phi^{(NSF)}$ is the same as \eqref{Thpot} with the energy, denoted now $E^{(NSF)}$,
being the energy entering \eqref{5flux}, the number of moles $N^{(NSF)}=\frac{1}{M_{mol}}\int d\rr \rho(\rr)$, where $M_{mol}$ is the molecular mass of the fluid under consideration, and  the entropy $S^{(NSF)}=\int s(\rr)$. The energy $E^{(NSF)}$ is assumed to be invariant with respect to the transformation $(\rho,\uu,s)\rightarrow (\rho,-\uu,s)$.

In order to investigate solutions to \eqref{5eqs} in more details, we can again take two routes. Either we follow example of \cite{VillaniV} and take  the  route of a direct and rigorous investigation
or we take an indirect route inspired by the methods of statistical mechanics. As in the previous sections, we  choose the latter. Assuming that the  forces driving fluids to equilibrium, denoted  $X^{(NSF)}$,  are proportional to the gradient of velocity (the Navier-Stokes force)  and the gradient of temperature (the Fourier force), introducing dissipation potential $\Xi^{(NSF)}$ that is a quadratic function of $X^{(NSF)}$,
and supplying the Euler hydrodynamic equations \eqref{5eqs} with the corresponding to it dissipative terms,  we arrive at the governing equations of the Navier-Stokes-Fourier fluid mechanics. It  can be shown (e.g. by following the Chapman-Enskog method) that solutions to the Navier-Stokes-Fourier equations approximate well asymptotic solutions to the Boltzmann equation \eqref{BE}. We can thus regard the Navier-Stokes-Fourier dissipation in the setting of fluid mechanics as an analog of the Boltzmann collision dissipation in the setting of kinetic theory and anticipate that the approach to the spatially homogeneous equilibrium in the Boltzmann kinetic theory proceeds in the same way also in the Navier-Stokes-Fourier fluid mechanics.

More specifically, we introduce two manifolds
\begin{equation}\label{MNSF}
\mathcal{M}_{NSF}=\{f\in M^{(hyd)}| X^{(NSF)}=0\}
\end{equation}
and
\begin{equation}\label{Minc}
\mathcal{M}_{inc}=\{f\in M^{(hyd)}| \rho=const., \,\,div\,\vv=0\}
\end{equation}
where $M^{(hyd)}$ denotes the state space of the classical hydrodynamics.
The manifold $\mathcal{M}_{NSF}$ is composed of hydrodynamic fields  for which the conjugates of $\uu(\rr)$ and $e(\rr)$ are spatially homogeneous but  the field $\rho(\rr)$  can still change with $\rr$. The manifold $\mathcal{M}_{NSF}$
plays in the Navier-Stokes-Fourier hydrodynamics an analogical role as the manifold $\mathcal{M}_{coll}$  (see \eqref{Mcoll})  plays in the Boltzmann kinetic theory. The manifold $\mathcal{M}_{inc}$ is the manifold on which the fluid has the spatially homogeneous mass density and $div\,\vv=0$.
The thermodynamic equilibrium state is the state for which the thermodynamic potential $\Phi^{(NSF)}$ reaches its minimum, it  is an element of both $\mathcal{M}_{NSF}$ and $\mathcal{M}_{inc}$.
We conjecture that the approach to the thermodynamic equilibrium  in the Navier-Stokes-Fourier hydrodynamics proceeds in the same way as in the Boltzmann kinetic theory (see the third paragraph after \eqref{Thpot}). We thus conjecture that solutions to the Navier-Stokes-Fourier hydrodynamic equations  \eqref{5eqs} approach rapidly  the vicinity of $\mathcal{M}_{NSF}$ and $\mathcal{M}_{inc}$  but never touch them (due to the influence of the Euler term - the right hand side of \eqref{5eqs} - that is analogical to the free flow term in \eqref{BE})  except at the end of the time evolution where all hydrodynamic fields are independent of $t$ and $\rr$.

Results reported in \cite{VillaniB}, \cite{Majda}, \cite{Lions},  \cite{Feireisl}, \cite{Karlin} and in \cite{Ewelina} are related to the conjecture.
The relation  to the properties of solution to the Boltzmann equation found in \cite{VillaniB} is based on the relation between solutions to  \eqref{5eqs} and asymptotic solutions to \eqref{BE} that can be established by following, for example,  the well known Chapman-Enskog type investigation of solutions to \eqref{BE}.
The authors of Refs.\cite{Majda},
\cite{Lions} have proven in  the case of isentropic flow and the authors of Ref. \cite{Feireisl} in the case of non-isentropic flows that in the low Mach number limit solutions to the  Navier-Stokes-Fourier equations  approach solutions to the incompressible  Navier-Stokes-Fourier equations (i.e. the Navier-Stokes-Fourier equations constrained to the manifold $\mathcal{M}_{inc}$). Thermodynamic interpretation of these result has been suggested in \cite{Karlin}.
The authors of Ref.\cite{Ewelina}  have  shown that, in the high Mach number limit,  solutions to the  Navier-Stokes-Fourier equations  approach solutions to a diffusion equation in the mass density $\rho(\rr)$ which then leads to $\rho=const$.

Now we turn to the Vlasov gas.  We do not expect the setting of the classical fluid mechanics recalled above to be suitable for describing its dynamics. Appropriate extensions of \eqref{fmsv}-\eqref{pp5} can however become suitable. What  shall we consider to be  the requirement for the suitability? We suggest it to be the  appearance of the Landau damping in  solutions to the governing equations. In the setting of kinetic theory, as we have seen in the previous sections,  the Landau damping is  the tendency to the spatial homogenization in the absence of the Boltzmann collision term.
We therefore define the \textit{Landau damping in fluid mechanics  as a tendency to the spatial homogenization in the absence of the Navier-Stokes-Fourier dissipation}. Our strategy in constructing the extensions will be the same as in kinetic theory. We select extra state variables characterizing the micro-turbulence internal structure, construct Hamiltonian time evolution equations governing their time evolution, introduce an appropriate dissipation of the extra state variables, and finally show that the dissipation induces the self-diffusion.

\subsection{Weakly nonlocal extension}\label{VHext}

The most straightforward way (we can say a "minimalist" way) to take into account  long range interactions in the setting of fluid mechanics is to use energy that is a nonlocal function of \eqref{fmsv}. With such energy, the role of the extra state variables is played by  higher  order gradients of the classical state variables \eqref{fmsv}.
Nonlocal and Hamiltonian extensions of the classical fluid mechanics have already been considered  in \cite{Grjsf} (in the context of fluids that are in the vicinity of gas-liquid phase transition) and in \cite{Van} (in the context of fluids involving self-diffusion). We shall therefore limit ourselves  only to pointing out the similarity  with Section \ref{3}.

If we use in the equation $\dot{A}=\{A,E\}^{(hyd)}; \forall A$, (see the text that follows \eqref{5br}) the  energy $E$ that depends on spatial derivatives of $\rho(\rr)$ and $\uu(\rr)$, we still obtain the time evolution equations \eqref{5eqs} but with modified fluxes. In particular, the mass flux $\JJ^{(\rho)}$ acquires an additional term involving spatial derivatives of $\uu(\rr)$. This new term is analogical to the second term in the first equation in \eqref{pVpsi}. Subsequent replacement of the   Navier-Stokes-Fourier  dissipation with a dissipation involving only higher  order spatial derivatives of $\uu(\rr)$) and an argument analogical to the argument of the absence of inertia that we made in the second equation in \eqref{pVpsi} leads then to the emergence of the diffusion term in the equation governing the time evolution of $\rho(\rr)$. We omit  details of the calculations since they can be found in  \cite{Grjsf} and in \cite{Van} and also since   the essence of the calculations and the arguments involved are  the same as those discussed in detail below in  the next extension in the context of the Reynolds-type extension.

\subsection{Reynolds-type extension}

In the second extension we follow the spirit of the Reynolds extension \cite{Reynolds} but the actual formulation of the extension is made by following Sections \ref{Reyext} and \ref{Vreg}.
We begin with two state variables
$x_1=(\rho_1(\rr),\uu_1(\rr),s_1(\rr))$ and  $x_2=(\rho_2(\rr),\uu_2(\rr),s_2(\rr))$ of the classical fluid mechanics. The kinematics of $(x_1,x_2)$ is expressed in the Poisson bracket $\{A,B\}^{(hyd12)}=\{A,B\}^{(hyd1)}+\{A,B\}^{(hyd2)}$, see e.g. \cite{PhysD} for a derivation.
Next,  we make one-to-one transformation to new state variables
$\xi=(\rho(\rr),\uu(\rr),s(\rr))$ and $\zeta=(\varrho(\rr), \nu(\rr), \varsigma(\rr))$: $\xi=x_1+x_2; \zeta=x_1$ (compare with \eqref{transf}). The fluid mechanics state variables $\xi$ characterize the overall behavior (similarly as the distribution function $\phi$ in Section \ref{Reyext}) and $\zeta$ the internal structure (similarly as $\psi$ in Section \ref{Reyext}). The Poisson bracket expressing kinematics of $(\xi,\zeta)$ becomes
\begin{equation}\label{brxz}
\{A,B\}^{Vhyd)}=\{A,B\}^{(\xi)}+\{A,B\}^{(\zeta)}+\{A,B\}^{(\xi\zeta)}
\end{equation}
where $\{A,B\}^{(\xi)}$ and $\{A,B\}^{(\zeta)}$ are the brackets \eqref{5br} and
\begin{eqnarray}\label{bbrxz}
\{A,B\}^{(\xi\zeta)}&=&\int d\rr\left[\varrho\left(\frac{\partial}{\partial r_{\alpha}}(A_{\rho})B_{\nu_{\alpha}}-\frac{\partial}{\partial r_{\alpha}}(B_{\rho})A_{\nu_{\alpha}}\right)\right.\nonumber \\
&&+\left.\varrho\left(\frac{\partial}{\partial r_{\alpha}}(A_{\varrho})B_{u_{\alpha}}-\frac{\partial}{\partial r_{\alpha}}(B_{\varrho})A_{u_{\alpha}}\right)\right.\nonumber \\
&&+\left.\varsigma\left(\frac{\partial}{\partial r_{\alpha}}(A_{s})B_{\nu_{\alpha}}-\frac{\partial}{\partial r_{\alpha}}(B_{s})A_{\nu_{\alpha}}\right)\right.\nonumber \\
&&+\left.\varsigma\left(\frac{\partial}{\partial r_{\alpha}}(A_{\varsigma})B_{u_{\alpha}}-\frac{\partial}{\partial r_{\alpha}}(B_{\varsigma})A_{u_{\alpha}}\right)\right.\nonumber \\
&&+\left.\nu_{\beta}\left(\frac{\partial}{\partial r_{\alpha}}(A_{u_{\beta}})B_{\nu_{\alpha}}-\frac{\partial}{\partial r_{\alpha}}(B_{u_{\beta}})A_{\nu_{\alpha}}\right)\right.\nonumber \\
&&+\left.\nu_{\beta}\left(\frac{\partial}{\partial r_{\alpha}}(A_{\nu_{\beta}})B_{u_{\alpha}}-\frac{\partial}{\partial r_{\alpha}}(B_{\nu_{\beta}})A_{u_{\alpha}}\right)\right]
\end{eqnarray}
The time evolution equations $\dot{A}=\{A,E\}^{(hyd)}; \forall A$, written explicitly, become
\begin{eqnarray}\label{5eqsext}
\frac{\partial \rho}{\partial t}&=& - \frac{\partial J^{(\rho)}_{\alpha}}{\partial r_{\alpha}}\nonumber \\
\frac{\partial u_i}{\partial t}&=& - \frac{\partial J^{(u)}_{i \alpha }}{\partial r_{\alpha}}          \nonumber \\
\frac{\partial s}{\partial t}&=&- \frac{\partial J^{(s)}_{\alpha}}{\partial r_{\alpha}}\nonumber \\
\frac{\partial \varrho}{\partial t}&=& - \frac{\partial}{\partial r_{\alpha}}\left(\varrho E_{\nu_{\alpha}}+\varrho E_{u_{\alpha}}\right)\nonumber \\
\frac{\partial \nu_i}{\partial t}&=& - \frac{\partial}{\partial r_{\alpha}}\left(\nu_i E_{\nu_{\alpha}}+\nu_iE_{u_{\alpha}}\right)\nonumber \\
&&-\varrho\frac{\partial E_{\varrho}}{\partial r_i}-\varsigma\frac{\partial E_{\varsigma}}{\partial r_i}-\nu_{\beta}\frac{\partial E_{\nu_{\beta}}}{\partial r_i}\nonumber \\
&&-\varrho\frac{\partial E_{\rho}}{\partial r_i}-\varsigma\frac{\partial E_s}{\partial r_i}-\nu_{\beta}\frac{\partial E_{u_{\beta}}}{\partial r_i}\nonumber \\
\frac{\partial \varsigma}{\partial t}&=&-\frac{\partial}{\partial r_{\alpha}}\left(\varsigma E_{\nu_{\alpha}}+\varsigma E_{u_{\alpha}}\right) \nonumber \\
\frac{\partial e}{\partial t}&=&- \frac{\partial J^{(e)}_{\alpha}}{\partial r_{\alpha}}\nonumber \\
\end{eqnarray}
with
\begin{eqnarray}\label{5fluxext}
\JJ^{(\rho)}&=&\rho E_{\uu}+\varrho E_{\nnu}\nonumber \\
\JJ^{(u)}&=&\ssigma+p\ddelta\nonumber \\
\JJ^{(s)}&=&s E_{\uu}+\varsigma E_{\nnu}\nonumber \\
\JJ^{(e)}&=&(e+p)E_{\uu}+\left(\varrho(E_{\varrho}+E_{\rho})+\varsigma(E_{\varsigma}+E_{s})+\nu_{\beta}(E_{\nu_{\beta}}+E_{u_{\beta}})\right)E_{\nnu}\nonumber \\
\end{eqnarray}
where
\begin{equation}\label{sig5ext}
\ssigma=\uu E_{\uu}+\nnu E_{\nnu}
\end{equation}
and
\begin{equation}\label{pp5ext}
p=-e+\rho E_{\rho}+sE_s+<\uu, E_{\uu}>+\varrho E_{\varrho}+\varsigma E_{\varsigma}+<\nnu, E_{\nnu}>
\end{equation}
The last equation in (\ref{5eqsext}) is a consequence  of the six equations above it but we have added it for the completeness.

We conjecture that this extended fluid mechanics is suitable for describing the time evolution of the Vlasov gas. In order to prove it, we can either investigate  solutions to \eqref{5eqsext} for an appropriately chosen energy $E(\rho,\uu,s,\varrho,\nnu,\varsigma)$ or follow the route that we have taken in the previous sections in the kinetic theory, namely the route on which we  make assumptions about the micro-turbulence (characterized now by  the fields  $\varrho(\rr), \nu(\rr), \varsigma(\rr)$), appropriately modify (regularize) \eqref{5eqsext}, and then investigate solutions of the regularized equations. As in the previous sections, we  take the latter route.

Before making the regularization we note that the general properties \eqref{BEfth} (or \eqref{BEfthV}, or \eqref{HthVV}) hold again for solutions to \eqref{5eqsext} with the energy $E$ appearing in \eqref{5fluxext}, entropy $S=\int d\rr s(\rr)$. and the number of moles $N=\frac{1}{M_{mol}}\int d\rr \rho(\rr)$. The energy $E$ is assumed to be invariant with respect to the transformation $(\rho,\uu,s,\varrho,\nnu,\varsigma)\rightarrow \rho,-\uu,s,\varrho,-\nnu,\varsigma)$.

We assume that the force driving the decay of the micro-turbulence is
\begin{equation}\label{XVm}
X^{(Vhyd)}_{i}=\nu^*_i
\end{equation}
and the dissipation potential
\begin{equation}\label{XiVhyd}
\Xi^{(Vhyd)}=\int d\rr X^{(Vhyd)}_{i}\frac{1}{2}\varrho\Lambda^{(Vhyd)}X^{(Vhyd)}_{i}
\end{equation}
where $\Lambda^{(Vhyd)}>0$ is a parameter.

With this dissipation, the equation  governing the time evolution of $s(\rr)$ becomes
\begin{equation}\label{sext}
\frac{\partial s}{\partial t}=- \frac{\partial J^{(s)}_{\alpha}}{\partial r_{\alpha}} +\sigma^{(Vhyd)}
\end{equation}
where the entropy production
\begin{equation}\label{sighyd}
\sigma^{(Vhyd)}= X^{(Vhyd)}_{ij}\varrho\Lambda^{(Vhyd)}X^{(Vhyd)}_{ij}>0
\end{equation}
and the
equation governing the time evolution of $\nnu(\rr)$ becomes
\begin{eqnarray}\label{dissnu}
\frac{\partial \nu_i}{\partial t}&=& - \frac{\partial}{\partial r_{\alpha}}\left(\nu_i E_{\nu_{\alpha}}+\nu_iE_{u_{\alpha}}\right)\nonumber \\
&&-\varrho\frac{\partial E_{\varrho}}{\partial r_i}-\varsigma\frac{\partial E_{\varsigma}}{\partial r_i}-\nu_{\beta}\frac{\partial E_{\nu_{\beta}}}{\partial r_i}\nonumber \\
&&-\varrho\frac{\partial E_{\rho}}{\partial r_i}-\varsigma\frac{\partial E_s}{\partial r_i}-\nu_{\beta}\frac{\partial E_{u_{\beta}}}{\partial r_i}\nonumber \\
&&-\varrho\Lambda^{(Vhyd)}\nu^*_i
\end{eqnarray}
We now assume that the energy $E$ and the parameter $\Lambda^{(Vhyd)}$ are chosen in such a way that in the later stage of the time evolution the dominant terms on the right hand side of \eqref{dissnu} are $-\varrho\Lambda^{(Vhyd)}\nu^*_i$ and    $-\varrho\frac{\partial \rho^*}{\partial r_i}$. Consequently, we replace \eqref{dissnu} by
\begin{equation}\label{111}
-\varrho\Lambda^{(Vhyd)}\nu^*_i-\varrho\frac{\partial \rho^*}{\partial r_i}=0
\end{equation}
If we now insert \eqref{111} into the first equation in \eqref{5eqsext} we arrive at the convection-diffusion equation
\begin{equation}\label{1112}
\frac{\partial \rho}{\partial t}=-\frac{\partial(Tu^*_i)}{\partial r_i}+\frac{\partial}{\partial r_i}\left(T(\Lambda^{(Vhyd)})^{-1}\frac{\partial \rho^*}{\partial r_i}\right)
\end{equation}
which manifestly displays the tendency towards  spatial homogenization.

Summing up, we have shown  that the setting provided by \eqref{5eqsext} predicts the Landau damping (as we have defined in fluid mechanics) and can be thus  suitable for describing the time evolution of the Vlasov gas.

\section{Concluding remarks}

Statistical mechanics has been very successful in elucidating  the passage from the microscopic viewpoint of macroscopic systems (i.e. the viewpoint that takes  into account  all the microscopic details) to the macroscopic viewpoint (in which all the information about the macroscopic systems comes from macroscopic measurements that ignore  unimportant microscopic details) by creating a multiscale hierarchy of mesoscopic levels  on which   varying amount of details is taken into account. The phenomena emerging in macroscopic observations (i.e. the phenomena that are not directly seen in the microscopic observations but that are nevertheless consequences of  the  microscopic dynamics) become less mysterious by seeing  them in the process of their  emergence. The archetype example is Boltzmann's illumination  of the emergence of the time irreversible approach to the  homogeneous in  space and Maxwellian in  velocities distribution of the Boltzmann gas by relating it to the ignorance of details of particle trajectories during collisions. An additional  understanding of the time irreversible approach is then gained by seeing it also on the level of fluid mechanics that is reduced from the Boltzmann kinetic theory.

In this paper we look at the emergence of the time irreversible approach to spatially homogeneous  distribution (called Landau damping) observed in the
Vlasov gas (a gas composed of particles interacting via a long range potential). The microscopic description is represented by the Vlasov kinetic equation. Although the  rigorous mathematical analysis of its solutions, developed in \cite{VillaniV}  does  reveal the Landau damping, its physical understanding is enhanced by presenting it on a new level, namely the level of  the extended kinetic theory. According to \cite{VillaniV}, Landau damping can be regarded as transfer of information from the spatial Fourier modes to the velocity Fourier modes. The regularization of the Vlasov equation consists of (in the nutshell): (i)   adoption  of the velocity Fourier modes as an extra independent state variable (interpreted as an internal  state variable characterizing the micro-turbulence), (ii) extension of the Vlasov equation to an equation governing the time evolution of the extended set of state variables (by keeping the Hamiltonian structure of the Vlasov equation), and (iii)  regularization of the extended Vlasov equation by introducing into it a friction-type decay of the micro-turbulence structure. Solutions to the extended and regularized Vlasov equation then show the Landau damping.
\\
\\
\\

\textbf{Acknowledgements}

We are grateful to Josef M\'alek and Milan Pokorn\'y for consulting the mathematical results in hydrodynamics.
This research was partially supported by  the Natural Sciences and Engineering
Research Council of Canada.
MP was supported by the Czech Science Foundation, project no.  17-15498Y.
\\
\\
\\

\end{document}